\documentclass[pra,twocolumn,titlepage,nofootinbib,amsmath,showpacs
]{revtex4}%
\usepackage{amsmath}
\usepackage{amsfonts}
\usepackage{amssymb}
\usepackage{graphicx}%
\begin{document}
\title{On the accuracy of the optical determination of the proton charge radius}
\author{Savely~G.~Karshenboim}
\email{savely.karshenboim@mpq.mpg.de}
\affiliation{Max-Planck-Institut f\"ur Quantenoptik, Garching, 85748, Germany}
\affiliation{Pulkovo Observatory, St.Petersburg, 196140, Russia}



\begin{abstract}
Determination of the proton charge radius by different methods has
produced an inconsistency. The most precise value (from spectroscopy of
muonic hydrogen) strongly disagrees with three less accurate values
(from spectroscopy of ordinary hydrogen and deuterium, from relative
measurements of the cross section of the elastic electron-proton scattering at
MAMI and from evaluation of the world data on absolute measurements
of $e-p$ cross sections). Here, we question the accuracy of the
determination of the proton charge radius by means of spectroscopy of
ordinary hydrogen and deuterium and demonstrate that its accuracy was
probably overestimated. In particular, we revisit determination from each
relevant transition and find that the results of two optical experiments,
which are the most statistically important, are not perfectly consistent. The
inconsistency is rather a `tension' between the results than their
discrepancy, however, it implies that a more conservative estimation of
the uncertainty is needed. With the more realistic estimation of the
uncertainty, the results for the proton charge radius from spectroscopy
of ordinary and muonic atoms are rather in fair agreement.
\end{abstract}
\maketitle

\section{Introduction}

\begin{figure}[thbp]
\begin{center}
\resizebox{1.0\columnwidth}{!}{\includegraphics{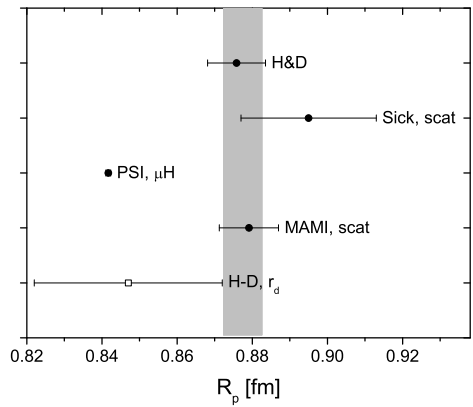}}
\end{center}
\caption{Determination of the rms proton charge radius.
The vertical belt is the CODATA 2010 recommended value \cite{codata2010}.
For details see \cite{my_adp}.
}
\label{fig:re}       
\end{figure}

For a few last years, since the publication in 2010 \cite{Nature},
the so-called proton-radius puzzle became a part of physics landscape. The
present-day statement is that the determination of the proton charge radius
from the Lamb shift in muonic hydrogen \cite{Science,efit} has an
uncertainty much smaller than that of the other determinations, but its
value is in a strong contradiction with them. The situation with
the proton rms charge radius $R_p$ is graphically presented in
Fig.~\ref{fig:re}, where the results are plotted following \cite{my_adp},
based on the CODATA analysis \cite{codata2010}.

The {\em H\/}\&{\em D\/} value is a result of the CODATA
evaluation \cite{codata2010} of spectroscopic data on 24 transitions
in atomic hydrogen and deuterium and it is roughly $5\,\sigma$ away
from the muonic {\em PSI\/} value \cite{Science}. Two more points are from
the elastic electron-proton scattering. The {\em Sick\/} result is from an
evaluation of world data on the absolute measurements of the cross
sections \cite{sick}, while the {\em MAMI\/} data point results from a
relative measurement of the cross sections at the Mainzer
Mikrotron (MAMI) \cite{mami}. The former is within $3\,\sigma$, while
the latter is within $5\,\sigma$ from the result from Paul Scherrer
Institut (PSI). Those three results, {\em H\/{\rm\&\/}D, MAMI {\rm and\/}
Sick's\/}, are apparently consistent. Averaging them, one finds the mean
CODATA value \cite{codata2010}  $7\,\sigma$ away from the muonic one,
which is indeed a strong discrepancy. There is also one more point from
combining the isotopic-shift measurement of the $1s-2s$ transition
in hydrogen and
deuterium with a result of the electron-deuteron scattering \cite{rdsick},
however, it is not conclusive.

This controversy is challenging, and various solutions have been proposed
including some unrealistic properties of the proton and unreasonable
behavior of higher-order QED effects as well as new physical phenomena.
No reasonable explanation within the standard physics has been suggested.

We believe that the solution lies in a different direction. The statement
on the discrepancy is valid only if we rely on the accuracy of the mentioned
results as they have been claimed. However, it often happens that the
accuracy is overestimated and here we revisit the accuracy of the
spectroscopic determination, which, as we mentioned, is within
$5\,\sigma$ from the muonic result.

We have revisited determination of the proton charge radius from
spectroscopy of ordinary hydrogen and deuterium. We basically follow the
evaluation in \cite{codata2010}. There 24 measurements of various
transitions in hydrogen and deuterium were utilized, which produced
22 partial values of the proton radius. Below we consider those values
in detail.

\section{Extraction of the Rydberg constant and the proton charge radius
from spectroscopy of hydrogen and deuterium}

The frequency of any transition in the hydrogen atom can be expressed in
terms of the Rydberg constant and the proton charge radius. The expressions
are very complicated, however, the dominant dependence on these two
parameters is very simple (see, e.g., \cite{my_rep})
\begin{eqnarray}\label{eq:ry}
hf(n'l'-nl)&=&-hcR_\infty\left(\frac{1}{n^2}-\frac{1}{{n'}^2}\right) \nonumber\\
&&+E_L(nl)-E_L(n'l')+...
\end{eqnarray}
where the Lamb shift of the $nl$ state includes dependence on the rms
proton charge radius
\begin{equation}
E_L(nl)=\frac{4\alpha^4mc^2}{3}\frac{m^2c^4 R_p^2}{\hbar^2}
\frac{\delta_{l0}}{n^3}+ E'_L(nl)\;.
\end{equation}

The remaining terms in (\ref{eq:ry}) and $E'_L(nl)$ indeed also depend
on $R_\infty$ and $R_p$. However, the required accuracy is relatively low
and does not interfere with any determination of the Rydberg constant and
the proton charge radius. Still, $E'_L(nl)$ contributes to the final
uncertainty of the determination of those constants
(see, e.g., \cite{codata2010}).

Many experimentally measured frequencies, such as those of $1s-2s$ or
$2s-8d_{5/2}$, depend both on the proton-radius term and the Rydberg-constant
term. Thus, one has to combine the two hydrogen frequencies to obtain a
value of the radius. On the contrary, certain experimental frequencies,
e.g., $f(2s-2p)$, do not include the Rydberg-constant term and the proton
radius can be found immediately. To obtain a value of the Rydberg constant
one has to combine such a transition with one, frequency of which contains
the Rydberg-constant term.


Deuterium data can be described in a similar way. Combining two deuterium
results one can find a value of the Rydberg constant and, applying it to a
hydrogen transition, one can extract a value of the proton radius.

Below we consider the hydrogen and deuterium data and their required
combinations in detail. The relevant spectroscopic data are summarized in
Table~\ref{t:HDfreq}. The frequencies listed there are `effective'
frequencies after introduction of the appropriate hyperfine-structure
correction to the originally measured transition frequencies. While
describing the data, we follow \cite{codata2010} (see Table XI there).

\begin{table}[htbp]
\begin{center}
\begin{tabular}{lcc}
Transition(s) & Frequency [kHz] & Ref. \\[0.8ex]
\hline
$f_{\rm H}({\rm 1s}-{\rm 2s})$& $ 2\,466\,061\,413\,187.080(34)$ &  \cite{mpq} \\[0.8ex]
$f_{\rm D}({\rm 1s}-{\rm 2s})-f_{\rm H}({\rm 1s}-{\rm 2s})$& $ 670\,994\,334.606(15)$  &  \cite{mpq_d} \\[0.8ex]
\hline
$f_{\rm H}({\rm 2s}-{\rm 8s})$ & $770\,649\,350\,012.0(8.6)$ &  \cite{paris1} \\[0.8ex]
$f_{\rm H}({\rm 2s}-{\rm 8d_{3/2}})$ & $ 770\,649\,504\,450.0(8.3)$ &  \cite{paris1} \\[0.8ex]
$f_{\rm H}({\rm 2s}-{\rm 8d_{5/2}})$ & $ 770\,649\,561\,584.2(6.4)$  &  \cite{paris1} \\[0.8ex]
$f_{\rm D}({\rm 2s}-{\rm 8s})$ & $ 770\,859\,041\,245.7(6.9)$  &  \cite{paris1} \\[0.8ex]
$f_{\rm D}({\rm 2s}-{\rm 8d_{3/2}})$ & $ 770\,859\,195\,701.8(6.3)$  &  \cite{paris1} \\[0.8ex]
$f_{\rm D}({\rm 2s}-{\rm 8d_{5/2}})$ & $ 770\,859\,252\,849.5(5.9)$  &  \cite{paris1} \\[0.8ex]
$f_{\rm H}({\rm 2s}-{\rm 12d_{3/2}})$ & $ 799\,191\,710\,472.7(9.4)$  &  \cite{paris2} \\[0.8ex]
$f_{\rm H}({\rm 2s}-{\rm 12d_{5/2}})$ & $ 799\,191\,727\,403.7(7.0)$  &  \cite{paris2} \\[0.8ex]
$f_{\rm D}({\rm 2s}-{\rm 12d_{3/2}})$ & $ 799\,409\,168\,038.0(8.6)$  &  \cite{paris2} \\[0.8ex]
$f_{\rm D}({\rm 2s}-{\rm 12d_{5/2}})$ & $ 799\,409\,184\,966.8(6.8)$  &  \cite{paris2} \\[0.8ex]
$f_{\rm H}({\rm 1s}-{\rm 3s})$ & $ 2\,922\,743\,278\,678(13)$   &  \cite{1s3s} \\[0.8ex]
\hline
$f_{\rm H}({\rm 2s}-{\rm 4s})- \frac{1}{4}f_{\rm H}({\rm 1s}-{\rm 2s})$ & $ 4\,797\,338(10)$ & \cite{obfMPQ} \\[0.8ex]
$f_{\rm H}({\rm 2s}-{\rm 4d_{5/2}}) - \frac{1}{4}f_{\rm H}({\rm 1s}-{\rm 2s})$ & $ 6\,490\,144(24)$  &  \cite{obfMPQ} \\[0.8ex]
$f_{\rm D}({\rm 2s}-{\rm 4s})- \frac{1}{4}f_{\rm D}({\rm 1s}-{\rm 2s})$ & $ 4\,801\,693(20)$  &  \cite{obfMPQ} \\[0.8ex]
$f_{\rm D}({\rm 2s}-{\rm 4d_{5/2}})- \frac{1}{4}f_{\rm D}({\rm 1s}-{\rm 2s})$ & $ 6\,494\,841(41)$  &  \cite{obfMPQ} \\[0.8ex]
$f_{\rm H}({\rm 2s}-{\rm 6s})- \frac{1}{4}f_{\rm H}({\rm 1s}-{\rm 3s})$ & $ 4\,197\,604(21)$  &  \cite{obfP} \\[0.8ex]
$f_{\rm H}({\rm 2s}-{\rm 6d_{5/2}}) - \frac{1}{4}f_{\rm H}({\rm 1s}-{\rm 3s})$ & $ 4\,699\,099(10)$  &  \cite{obfP} \\[0.8ex]
$f_{\rm H}({\rm 2s}-{\rm 4p_{1/2}})- \frac{1}{4}f_{\rm H}({\rm 1s}-{\rm 2s})$& $ 4\,664\,269(15)$  &  \cite{obfY} \\[0.8ex]
$f_{\rm H}({\rm 2s}-{\rm 4p_{3/2}}) - \frac{1}{4}f_{\rm H}({\rm 1s}-{\rm 2s})$ & $ 6\,035\,373(10)$  &  \cite{obfY} \\[0.8ex]
\hline
$f_{\rm H}({\rm 2s}-{\rm 2p_{3/2}})$ & $ 9\,911\,200(12)$  &  \cite{Hagley} \\[0.8ex]
$f_{\rm H}({\rm 2p_{1/2}}-{\rm 2s})$ & $ 1\,057\,845.0(9.0)$  &  \cite{Lundeen81} \\[0.8ex]
$f_{\rm H}({\rm 2p_{1/2}}-{\rm 2s})$ & $ 1\,057\,862(20)$  &  \cite{Newton} \\[0.8ex]
\end{tabular}
\caption{The H and D transition frequencies relevant for determination
of the Rydberg constant and the proton charge radius following Table XI
of \cite{codata2010}. Those 24 transition frequencies include 2 anchor
$1s-2s$ transitions, 11 other absolutely measured transitions, 8
combinations of optical transitions (`beat frequencies') and 3 microwave
frequencies.
\label{t:HDfreq}}
 \end{center}
 \end{table}

There are a few important components of the evaluation of the hydrogen
and deuterium spectroscopic data. At first, we have to mention the results
on the $1s-2s$ transition frequencies in hydrogen and deuterium measured
at MPQ (Max-Planck-Institut f\"ur Quantenoptik)
\cite{mpq,mpq_d}\footnote{Following \cite{codata2010}, we are to ignore
more recent measurements of the $1s-2s$ transition \cite{newopt}, which
have a superficial accuracy for our purposes.}. These measurements are
much more accurate than all other optical measurements (see
Table~\ref{t:HDfreq}) and they are to be used as their `companions' to
find the values of the proton radius (from the H transitions) or the
value of the Rydberg constant (from the D transitions). In the latter case,
we use the $1s-2s$ result in hydrogen to transform a value of $R_\infty$,
obtained from D transitions, into a value of $R_p$. Such combinations are
very helpful since the accuracy of the $1s-2s$ frequencies is better than
of any other optical transitions by more than an order of magnitude. We may
neglect any correlations between the extracted values of $R_p$ due to use
of data on the $1s-2s$ transitions.

Let's consider the relevant data in more detail. We do not follow the
historic chronological way and list the data according to their relevance.
As well as in \cite{codata2010}, we consider 24 transition frequencies.
Two of them, namely, $1s-2s$ in hydrogen and deuterium, serve as the `anchor'
transitions in our evaluation. They are used as the auxiliary companions
to each of the other 22 transition frequencies to find $R_\infty$, $R_p$ and
$R_d$. The most important among those 22 transition frequencies are the
optical data. All relevant absolute frequency measurements, except for the
anchor measurements of the $1s-2s$ transition, were performed at LKB
(Laboratoire Kastler Brossel). Three experiments include a measurement of
six $2s-8s/d$  \cite{paris1}, four $2s-12d$  \cite{paris2} and one
$1s-3s$ transition \cite{1s3s}. Except for the latter, the other
measurements were performed on both hydrogen and deuterium and they are
the most statistically important because of their accuracy (see below).

To take advantage of a possibility to extract $R_\infty$ by combining the
$1s-2s$ frequency with the other one, e.g., $2s-nl$, one has to be able
either to calculate $E_L^\prime$ for various levels or to find a relation
between them. An appropriate relation is based on a possibility to
obtain the difference \cite{my_rep,del1} (see also \cite{del:also})
\begin{equation}
\Delta_L(n)=E_L(1s)-n^3E_L(ns)=E_L^\prime(1s)-n^3E_L^\prime(ns)
\end{equation}
for various $ns$ states, the most important of which is the $2s$ state.
Meantime, for $l\neq0$ a calculation of the $E_L^\prime(nl)$ is not a problem.
That allows us to express all the transitions in terms of $R_\infty$ and
$E_L(2s)$, while all the other contributions are under control and do not
require accurate knowledge either of $R_\infty$ or of $R_p$. To extract $R_p$
from $E_L(2s)$ one has to deal with a complete theory of the Lamb shift
in hydrogen. In the CODATA evaluation \cite{codata2010} the expression
for $\Delta_L(n)$ has not been used explicitly, but their suggestions
on the character of the $n$-dependence of various contributions are
roughly equivalent to this approach.
Eventually, we follow the summary \cite{codata2010} of a relevant theoretical
contributions to the energy levels in H and D atoms (see also \cite{NIST:HD}).

Note that the deuterium values of $R_p$ involve absolutely the same theory
as hydrogenic ones. One can have in mind a picture, in which we first find
$R_d$ and next, applying the result on the isotopic shift of the $1s-2s$
transition as a constraint, obtain $R_p$. In this way we need a theory of
the deuterium Lamb shift and of the isotopic shift and such theories are in
general more complicated and less reliable than the one for the hydrogen
atom. The issue is the deuteron polarizability, which may involve additional
uncertainty. However, the contribution of the deuteron polarizability
eventually cancels out for $R_p$. That is easy to understand if we choose
a different prescription to utilize the same data. We use the isotopic shift
to find the $1s-2s$ frequency in D, next we combine it with a deuterium
transition of interest to find a value of the Rydberg constant. All that
does not require any theory for the deuteron polarizability (since it has
a trivial state dependence being proportional to $\delta_{l0}/n^3$ and does
not contribute to $\Delta_L(n)$ for deuterium). With the value of the
Rydberg constant extracted from deuterium, we obtain a value of the
proton radius from the $1s-2s$ transition in hydrogen. All that requires
a theory of the Lamb shift in hydrogen, but not in deuterium.

\section{The data and the results}

\begin{figure*}[thbp]
\begin{center}
\resizebox{0.9\textwidth}{!}{\includegraphics{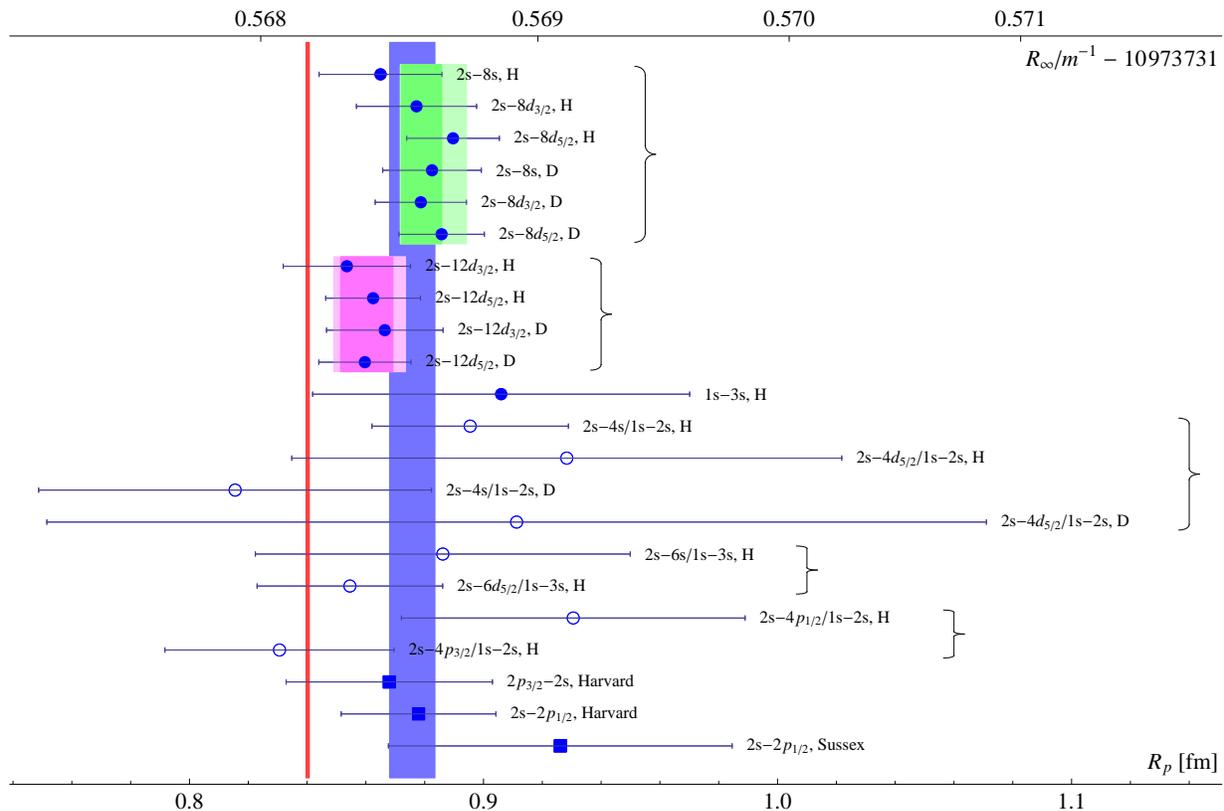}}
\end{center}
\caption{Determination of the proton charge radius and the Rydberg constant
from various H and D transitions. Each of 22 partial values of the proton
radius and the Rydberg constant is obtained
by using a certain unique measurement of a transition frequency or their
combination (the transition is used as the label) and the frequency of
one or two anchor $1s-2s$ transitions as explained in the text. The
transitions are grouped by experiments; the related references could be
found in Table~\ref{t:HDfreq}. The closed circles are for 11 absolute optical
frequency measurements, 8 open ones are for the differential optical
frequency measurements, and 3 squares are for the microwave transitions.
The long blue broad belt is for the CODATA-2010 all-spectroscopic value
(see adjustment 6 in Table XXXVIII of \cite{codata2010}).
The long narrow red belt (which looks as a bold line) is for the value from
the muonic hydrogen Lamb shift \cite{Science,efit}. The other filled areas
are for the average value of two crucial
experiments \cite{paris1} and \cite{paris2},
performed at LKB. The `dark' part of each area is for the mean value
without taking any correlations into account while the `light' one is
with taking them into account following \cite{paris2}.}
\label{f:overall}       
\end{figure*}

The relevant transitions are summarized in Table~\ref{t:HDfreq} and
Fig.~\ref{f:overall}.  We present in Fig.~\ref{f:overall} the partial
results of a determination of the proton charge radius and the Rydberg
constant. The partial results
on determination of the deuteron radius look
similar, because $R_d$, and $R_p, R_\infty$ are strongly correlated to each
other. As we mentioned, the most accurate results on the proton charge
radius come from a comparison of 11 optical transitions
\cite{paris1,paris2,1s3s} with the anchor data on the $1s-2s$ transitions.
The other experiments \cite{obfMPQ,obfP,obfY} and
\cite{Hagley,Lundeen81,Newton} produce frequencies which do not include
the Rydberg term and allow us to determine $R_p$ directly, However, they are
statistically substantially less important as clearly seen
in Fig.~\ref{f:overall} where we summarize all the 22 determinations.
To determine the Rydberg constant from the experiments, which produce a value
of the proton radius without involving of the anchor data, one has to combine
the derived proton radius with the $1s-2s$ transition frequency.

Various experimental correlations are presented and the extracted values
of the Rydberg constant are as [in]dependent from  each other as the data
are. The additional correlations due to the evaluation procedure are
marginal for the Rydberg constant. Besides, for the proton radius there is
an additional (to the determination of the Rydberg constant) theoretical
correlation due to the application of the same value of $E'_L(2s)$ to all
the evaluations. This may shift all the values of the proton radius,
extracted from H \& D spectroscopy, the same way but does not
affect their scatter.

The extracted partial values in Fig.~\ref{f:overall} are strongly
correlated not only because some use similar methods and perform
measurements at a similar set up, but also because those 22 transition
frequencies were measured in 9 experiments only and, in particular, ten
most statistically significant results originate from two experiments
\cite{paris1,paris2} performed at essentially the same set up.

While $\chi^2$ of the distribution around the average value \cite{codata2010}
seems reasonable, we note that the distribution of the results in
Fig.~\ref{f:overall} around the mean value is not random. A comparison of
the two most accurate experiments \cite{paris1} and \cite{paris2} shows
that there is a certain systematic effect which has been missed in the
evaluation of data and which is responsible for the difference between
these two groups of the results. While most of $2s-8s/d$ transitions
strongly disagree with the muonic result (red line), the $2s-12d$ data are
rather in fair agreement with it. The controversy between the results of
two LKB measurements is explicitly presented in their summary table for
the Rydberg constant (see the results obtained using the `$1/n^3$
scaling law' in Table II in \cite{paris2}, see also \cite{epjparis}
for detail), but it is not discussed there.

\begin{figure}[thbp]
\begin{center}
\resizebox{1.0\columnwidth}{!}{\includegraphics{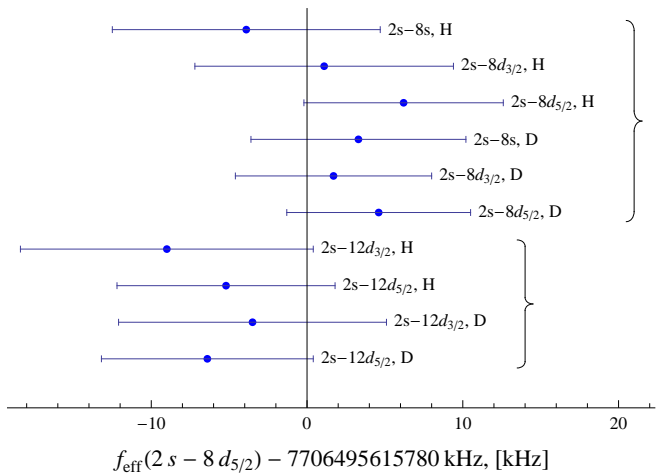}}
\end{center}
\caption{The effective frequency of the $2s-8d_{5/2}$ transition in the
hydrogen atom calculated from various $2s-8s/8d/12d$ H and D transitions
from \cite{paris1} and \cite{paris2}. The theoretical uncertainty of
the re-evaluation is negligible, being below 1 kHz (cf. \cite{NIST:HD,epjparis}).}
\label{f:nist10}       
\end{figure}

We have to comment on the theory applied to produce values of $R_p$ in
Fig.~\ref{f:overall} from the $2s-8s/8d/12d$ transitions. As we have
explained, we use for the extraction basically the same theory, a theory
of the Lamb shift in hydrogen. There is a small difference in theoretical
expressions because we studied different $2s-nl$ transitions. Since $n$
is rather high ($n=8,12$), we find that details of the theory of the higher
excited states $8s, 8d, 12d$ are rather of marginal importance for the
accuracy of the extraction. In other words, a small inconsistency, which shows
up in the extraction, already exists in the experimental data of two
experiments \cite{paris1} and \cite{paris2}, since all the transitions
are evaluated in approximately the same way. In particular, one can find
an effective value of the frequency for the $2s-8d_{5/2}$ transition in
hydrogen from each of ten H or D transitions under question. The results
are plotted in Fig.~\ref{f:nist10}. The QED computational uncertainty is
substantially below 1 kHz as well as the standard uncertainty (from H \& D
spectroscopy) due to applied values of the Rydberg constant and the proton
radius (cf. \cite{NIST:HD,epjparis}). (If one uses a value of the Rydberg
constant consistent with the $\mu$H Lamb shift to find an effective value
of the $2s-8d_{5/2}$ frequency from a $2s-12d$ transition, such an effective
value is to be higher by roughly 1 kHz, which does not remove the systematic
pattern).

We expect the value of the proton radius to lie between 0.85 and 0.89 fm.
Until the nature of the systematic effect is clarified, we see no conclusive
reasons to expect that the spectroscopic value of the proton radius is more
accurate. Indeed, the suggested constraint still does not agree literarily
with the muonic-hydrogen value, however, the discrepancy is not statistically
significant. It is also important that two LKB experiments are correlated.
It is possible that both series are affected by the same systematic effect
and the true value of $R_p$ may lie outside of the interval 0.85--0.89 fm.

The systematic pattern in the distribution of the individual results in Figs.~\ref{f:overall} and~\ref{f:nist10} has not drawn attention previously in part because the overall
procedure has a good $\chi^2$ value, thanks to the less accurate
determinations and thanks to the fact that the ten questionable transitions produce a non-random distribution with all the data being roughly one sigma above \cite{paris1} or below \cite{paris2} the average value. We also note, that the attention was for various reasons concentrated
on internal consistency of the pure hydrogenic data (cf., e.g., \cite{paris2})
As for the hydrogen values by themselves, the difference between the
results of \cite{paris1} and \cite{paris2} is not that statistically
significant.


Concluding, the distribution of the data for the LKB experiments
\cite{paris1} and \cite{paris2} shows perfect internal agreement between different
measurements within each group and a small, but clear difference between the two
groups of the data, apparently caused by a certain systematic effect. The effect
should be understood before making any statement about a contradiction
between the results on the proton radius from spectroscopy of muonic
and ordinary atoms.

The question of an expansion of the uncertainty for the derived value of the proton radius from the H \& D spectroscopy might be considered as disputable. One may in principle object our estimation of the required expansion, but not the presence of a systematic effect by itself. The latter must require a certain extension of the uncertainty. In spite of the apparent presence of a systematic effect, it would be helpful if the LKB experiments will be re-done in one or other way, in order to check whether their results are reproducible. (Indeed, a competitive independent measurements of any transition in hydrogen and deuterium would be also greatly appreciated to resolve the problem.)


\begin{figure}[thbp]
\begin{center}
\resizebox{1.0\columnwidth}{!}{\includegraphics{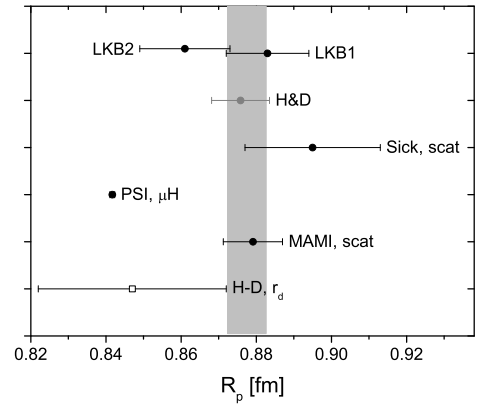}}
\end{center}
\caption{Determination of the rms proton charge radius. We follow
Fig.~\ref{fig:re}. The former value from hydrogen and deuterium spectroscopy
is shown as a grey point. Two additional points, labeled as LKB1 and LKB2,
are related to the evaluation of two LKB experiments \cite{paris1,paris2},
discussed above, taking into account their correlation following \cite{paris2}.
}
\label{fig:re:new}       
\end{figure}

The result of the systematic effect is not a `contradiction' between
different parts of the optical data. We should rather speak about a
`tension' between the data. However, it crucially affects the accuracy of
any overall statistical evaluation of the data. Actually, it rather prevents
it at a desirable level of accuracy. The level of accuracy below one
percent for the radius cannot be claimed until the effect is understood.
The present situation on this issue is summarized in
Fig.~\ref{fig:re:new}.

With such an extended understanding of the uncertainty of the
spectroscopic value, the contradiction is now between the muonic result
and the elastic-scattering results \cite{mami,sick} only.
Those scattering results
have been obtained by somewhat different techniques, and, in particular,
the MAMI cross sections were measured in arbitrary units in \cite{mami},
while the world data, evaluated in \cite{sick}, were obtained from absolute
measurements. Nevertheless, the fitting procedure has various features
in common. We cannot claim any longer that the muonic value disagrees with
results, obtained by a few independent methods. The discrepancy in the
determination of the proton radius might in principle be attributed to
underestimation of the systematic uncertainty due to the fitting procedure.

The author is grateful to S.I. Eidelman, V.G. Ivanov, E.Yu. Korzinin,
R. Pohl, F. Nez, and Th. Udem for useful discussions.


\end{document}